\documentstyle[12pt]{article}

\begin{document}

\setcounter{page}{1}

\pagestyle{plain} \vspace{1cm}
\begin{center}
\Large{\bf Thermodynamics of an Evaporating Schwarzschild Black Hole in
Noncommutative Space}\\
\small \vspace{1cm}
{\bf Kourosh Nozari}\quad \quad and \quad \quad {\bf Behnaz Fazlpour }\\
\vspace{0.5cm} {\it Department of Physics,
Faculty of Basic Science,\\
University of Mazandaran,\\
P. O. Box 47416-1467,
Babolsar, IRAN\\
e-mail: knozari@umz.ac.ir}
\end{center}
\vspace{1.5cm}
\begin{abstract}
We investigate the effects of space noncommutativity and the
generalized uncertainty principle on the thermodynamics of a
radiating Schwarzschild black hole. We show that evaporation process
is in such a way that black hole reaches to a maximum temperature
before its final stage of evolution and then cools down to a
nonsingular remnant with zero temperature and entropy. We compare
our results with more reliable results of string theory. This
comparison Shows that GUP and space noncommutativity are similar
concepts at least from the view point of black hole thermodynamics.

\end{abstract}
\newpage
\section{Introduction}
After thirty years of intensive research in the field of radiating
black holes[1], various aspects of the problem still remain under
debate. For example the last stage of black hole evaporation is not
obvious in some respects. The string/black hole correspondence
principle [2] suggests that in this extreme regime stringy effects
cannot be neglected. In spite of the promising results that string
theory has had in quantizing gravity, the actual calculations of the
Hawking radiation are currently obtained by means of quantum field
theory in curved space [3]. In fact the black hole evaporation
occurs in a semiclassical regime, namely when the density of
gravitons is lower than that of the matter field quanta.
Nevertheless, the divergent behavior of the black hole temperature
in the final stage of the evaporation remains
rather obscure.\\
In addition to string theory itself, which provides an elegant
framework for incorporation of quantum gravity effects in black hole
physics by direct state counting, several alternative approaches to
incorporate quantum gravity effects in the calculation of black hole
thermodynamics have been proposed. These approaches can be
classified as follows:
\begin{itemize}
\item
Generalized Uncertainty Principle(GUP)\\
Existence of a nonzero minimal length scale (which leads to finite
resolution of spacetime structure) can be addressed in GUP(see[4]
and references therein). From a heuristic argument, one can use GUP
to find modification of Bekenstein-Hawking formalism of black hole
thermodynamics[5,6,7,8]. The
main consequences of this approach are summarized as follows:\\
Black hole evaporation ends up with a phase consisting a remnant
with zero entropy and there exists a finite temperature that black
hole can reach in its final stage of evaporation. This picture
differs drastically with Bekenstein-Hawking prescription which
accepts the total evaporation of Black holes.
\item
Modified Dispersion Relations(MDRs)\\
MDRs induced modification of black hole thermodynamics have their
origin on loop quantum gravity considerations(MDRs are signature of
Lorentz invariance violation at high energy sector of the field
theory). Attempts to modify Bekenstein-Hawking formalism based on
MDRs show more or less the same behaviors as GUP framework, but now
we find severe constraints on the functional form of MDRs when we
compare our results with string theory more reliable results[9,10].
\item
Noncommutative Geometry\\
Noncommutativity eliminates point-like structures in favor of
smeared objects in flat spacetime. Based on this idea, several
attempts have been performed to find modification of
Bekenstein-Hawking formalism of black hole thermodynamics within
noncommutative geometry[11,12]. The consequences of
these attempts are as follows:\\
The end-point of black hole evaporation is a zero temperature
extremal remnant with no curvature singularity.
\end{itemize}
In this paper we are going to proceed one more step in the line of
third alternative i.e.  Noncommutative Geometry. Our strategy
differs with existing literatures in two main respects: we don't
consider smeared picture of objects in noncommutative spacetime(as
has been considered in [11,12]), instead we deal with coordinate
noncommutativity which results modification of Schwarzschild radius.
Also we consider possible generalization of uncertainty principle
within a string theory point of view. We calculate entropy-area
relation and compare our results with more reliable results of
string theory(calculated based on direct state-counting).
This comparison shows that GUP and space noncommutativity are essentially similar concepts.\\
In which follows we suppose $c=\hbar=G=1$.

\section{ Black Hole Thermodynamics in GUP Framework }
The canonical commutation relations in a commutative spacetime
manifold are given as follows
\begin{equation}
\left[ { x}_i,{ x}_j \right]=0,\;\;\; \left[ { x}_i,{p}_j \right]=i
\delta_{ij},\;\;\; \left[ { p}_i,{ p}_j \right]=0.
\end{equation}
From a string theory point of view, existence of a minimal length
scale can be addressed in the following generalized uncertainty
principle
\begin{equation}
\delta x\delta p \geq \frac{1}{2} \bigg(1+\beta (\delta
p)^2+\gamma\bigg),
\end{equation}
where $\beta$ is string theory parameter related to minimal length.
Since we are dealing with absolutely minimum position uncertainty we
set $\;\gamma=\beta {\langle p\rangle^{2}}$ and therefore the
corresponding canonical commutation relation becomes
\begin{equation}
[x,p]=i(1+\beta p^{2}).
\end{equation}
The canonical commutation relations in commutative spacetime with
GUP become
\begin{equation}
\left[ {x}_i,{ x}_j \right]=0,\;\;\; \left[ { x}_i,{ p}_j \right]=i
\delta_{ij}(1+\beta { p^{2}} ),\;\;\; \left[ { p}_i,{ p}_j
\right]=0.
\end{equation}
Now consider the geometry of Schwarzschild spacetime with the
following metric
\begin{equation}
 ds^2= f(r) dt^2 -
\frac{dr^2}{f(r)}- r^2 \left( d \theta^2 +\sin^2 \theta d\phi^2
\right),
\end{equation}
where $f(r)= 1 - \frac{2M}{r}$. There is a horizon at $r_{\rm s}=2M$
with the following area
\begin{equation}
A=r_{\rm s}^2 \int_{0}^{2\pi} d \phi \int_{0}^{\pi} \sin \theta d
\theta= 4 \pi r_{\rm s}^2=16 \pi M^2.
\end{equation}
Bekenstein-Hawking formalism of black hole thermodynamics gives the
following relations for temperature and entropy of black hole
\begin{equation}
T_{\rm H}=\frac{1}{8 \pi M}
\end{equation}
and
\begin{equation}
S=4\pi M^{2}
\end{equation}
respectively. Within GUP framework, these equations should be
modified to incorporate quantum gravity effects. We use Bekenstein's
argument type considerations to find GUP induced modification of
black hole thermodynamics. For simplicity, consider the following
GUP
\begin{equation}
\delta x\delta p \geq \frac{1}{2} \bigg(1+\beta (\delta p)^2\bigg).
\end{equation}
A simple calculation gives,
\begin{equation}
\delta p\simeq\frac{\delta
x}{\beta}\Big[1\pm\sqrt{1-\frac{\beta}{(\delta x)^2}}\Big].
\end{equation}
Here to achieve correct limiting result we should consider the minus
sign in round bracket. In original Bekenstein approach, from a
heuristic argument based on Heisenberg uncertainty relation, one
deduces the following equation for Hawking temperature of black
hole,
\begin{equation}
T_H\approx \frac{\delta p}{2\pi}.
\end{equation}
Therefore, in the framework of generalized uncertainty principle,
modified black hole temperature is as follows
\begin{equation}
T^{GUP}_{H}\approx \frac{\delta
x}{2\pi\beta}\Big[1-\sqrt{1-\frac{\beta}{(\delta x)^2}}\Big].
\end{equation}
Within black hole near horizon geometry, since $\delta x\sim r_s$
where $r_s=2M$, one can write this equation in such a way that can
be comparable with equation (7):
\begin{equation}
T^{GUP}_{H}\approx
\frac{M}{\pi\beta}\Big[1-\sqrt{1-\frac{\beta}{4M^2}}\Big].
\end{equation}
which leads to the following relation
\begin{equation}
T^{GUP}_{H}\approx \frac{1}{8\pi M}\Big[1+\frac{\beta}{16
M^2}+\frac{\beta^2}{128  M^4}\Big],
\end{equation}
up to second order in $\beta$. Obviously, when quantum gravitational
effects are negligible, that is when $\beta\rightarrow 0$, this
relation gives (7) as a manifestation
of correspondence principle.\\
Now consider a quantum particle that starts out in the vicinity of
an event horizon and then ultimately absorbed by black hole. For a
black hole absorbing such a particle with energy $E$ and size $R$,
the minimal increase in the horizon area can be expressed as
\begin{equation}
(\Delta A)_{min}\geq 4(\ln2)ER,
\end{equation}
then one can write
\begin{equation}
(\Delta A)_{min}\geq 8 (\ln2)\delta p \delta x,
\end{equation}
where $E\sim c\delta p $ (with $c=1$) and  $R\sim 2\delta x$.
Using equation (10) for $\delta p$, we find
\begin{equation}
(\Delta A)_{min}\simeq\frac{2(\ln2) A}{\beta
\pi}\Big[1-\sqrt{1-\frac{4\pi\beta}{A}}\Big]
\end{equation}
where we have defined $A=4\pi (\delta x)^2$. Now we should determine
$\delta x$. Since our goal is to compute microcanonical entropy of a
large black hole, near-horizon geometry considerations suggests the
use of inverse surface gravity or simply the Schwarzschild radius
for $\delta x$. Therefore, $\delta x\sim r_s$ and defining $4\pi
r_s^2=A$ and $(\Delta S)_{min}=b=constant$, then it is easy to show
that,
\begin{equation}
\frac{dS}{dA}\simeq\frac{(\Delta S)_{min}}{(\Delta A)_{min}}\simeq
\frac{b \beta\pi}{2
(\ln2)A\Big[1-\sqrt{1-\frac{4\pi\beta}{A}}\Big]}.
\end{equation}
Note that $b$ can be considered as one bit of information since
entropy is an extensive quantity. Considering "calibration factor"
of Bekenstein as $\ln2$, the minimum increase of entropy(i.e. $b$),
should be $\ln 2$. Now we should perform integration. There are two
possible choices for lower limit of integration, $A=0$ and $A=A_p$ .
Existence of a minimal observable length leads to existence of a
minimum event horizon area, $A_p = 4\pi(\delta x_{min})^{2}$. So it
is physically reasonable to set $A_p$ as lower limit of integration.
Based on these arguments, we can write
\begin{equation}
S\simeq\int_{A_p}^A \frac{\beta\pi}{2
A\Big[1-\sqrt{1-\frac{4\pi\beta}{A}}\Big]}dA.
\end{equation}
An integration gives
\begin{equation}
S\simeq\frac{A}{4}-\frac{\pi\beta}{4}\ln {\frac
{A}{4}}+\sum_{n=1}^\infty c_{n}\Big(\frac{4}{A}\Big)^{n}+\cal{C},
\end{equation}
where $\cal{C}$ is a constant. This is an interesting result which
shows the logarithmic leading order correction plus a power series
expansion in terms of inverse of area. Up to third order in
$\frac{1}{A}$, we find
\begin{equation}
S\simeq\frac{A}{4}-\frac{\pi\beta}{4}\ln {\frac {A}{4}}
+\Big(\frac{\pi\beta}{4}\Big)^{2} \Big(\frac{4}{A}\Big)+
 \Big(\frac{\pi\beta}{4}\Big)^{3} \Big(\frac{4}{A}\Big)^{2}
 -3\Big(\frac{\pi\beta}{4}\Big)^{4}\Big(\frac{4}{A}\Big)^{3}+\cal{C},
\end{equation}
where
\begin{equation}
{\cal{C}}=-\frac{A_p}{4}+\frac{\pi\beta}{4}\ln {\frac {A_p}{4}}
-\Big(\frac{\pi\beta}{4}\Big)^{2} \Big(\frac{4}{A_p}\Big)-
 \Big(\frac{\pi\beta}{4}\Big)^{3} \Big(\frac{4}{A_p}\Big)^{2}
 +3\Big(\frac{\pi\beta}{4}\Big)^{4}\Big(\frac{4}{A_p}\Big)^{3}.
\end{equation}
It is obvious that when $A=A_{p}$, $S\rightarrow0$ and therefore
black hole remnant should have zero entropy. A result which is
physically acceptable since small classical fluctuations are not
allowed at remnant scales because of existence of minimal observable
length.

\section{Black Hole Thermodynamics in Noncommutative Geometry}
A noncommutative space can be realized by the coordinate operators
satisfying
\begin{equation}
\left[ {\hat x}_i,{\hat x}_j \right] = i
\theta_{ij},\;\;\;i,j=1,2,3,
\end{equation}
where $\hat x$'s are the coordinate operators and $\theta_{ij}$ is
the noncommutativity parameter with dimension of $(length)^2$.
Canonical commutation relations in noncommutative spaces read
\begin{equation}
\left[ {\hat x}_i,{\hat x}_j \right]=i \theta_{ij},\;\;\; \left[
{\hat x}_i,{\hat p}_j \right]=i \delta_{ij},\;\;\; \left[ {\hat
p}_i,{\hat p}_j \right]=0,
\end{equation}
Now, we note that there is a new coordinate system
\begin{equation}
\ x_i={\hat x}_i + \frac{1}{2} \theta_{ij} {\hat p}_j,\;\;\;
p_i={\hat p}_i
\end{equation}
with these new variables, $x_i$'s satisfy the usual(commutative)
commutation relations
\begin{equation}
\ [x_i,x_j] = 0,\;\;\;[x_i,p_j] = i\delta_{ij},\;\;\;[p_i,p_j] = 0.
\end{equation}
Note that noncommutativity is an intrinsic characteristic
of underlying manifold.\\
For a noncommutative Schwarzschild black hole, we have[13,14]
\begin{equation}
f(r)=\Big( 1 - \frac{2M}{\sqrt{{\hat r}{\hat r}}}\Big) ,
\end{equation}
where ${\hat r}$ satisfies (25). The horizon of the noncommutative
Schwarzschild metric as usual satisfies the condition ${\hat
g}_{00}=0$ which leads to
\begin{equation}
1 - \frac{2M}{\sqrt{{\hat r}{\hat r}}}=0 .
\end{equation}
If in this relation we change the variables ${\hat x}_i$ to $x_i$,
and then using(25), the horizon of the noncommutative Schwarzschild
black hole satisfies the following condition
\begin{equation}
1 -
\frac{2M}{\sqrt{(x_i-\frac{\theta_{ij}p_j}{2})(x_i-\frac{\theta_{ik}p_k}{2})}}=0.
\end{equation}
This leads us to the following relation
\begin{equation}
1-\frac{2M}{r} \left( 1 + \frac{x_i \theta_{ij} p_j}{2r^2} -
\frac{\theta_{ij} \theta_{ik} p_j p_k}{8r^2} \right) + {\cal O}
(\theta^3) +...=0,
\end{equation}
where $\theta_{ij}=\frac{1}{2} \epsilon_{ijk} \theta_k$. Using the
identity $ \epsilon_{ijr} \epsilon_{iks}= \delta_{jk} \delta_{rs} -
\delta_{js} \delta_{rk}$, one can rewrite (30) as follows
\begin{equation}
1 - \frac{2M}{r} - \frac{M}{2 r^3} \left[ {\vec L}.{\vec \theta} -
\frac{1}{8 } \left( p^2 \theta^2 -({\vec p}.{\vec \theta})^2 \right)
\right] + {\cal O}(\theta^3)+...=0,
\end{equation}
where $L_k=\epsilon_{ijk} x_i p_j$,  $p^2={\vec p}.{\vec p}\,$  and
$\,\theta^2={\vec \theta}.{\vec \theta}$ . If we set
$\theta_3=\theta$ and assuming that remaining components of $\theta$
all vanish (which can be done by a rotation or a re-definition of
the coordinates), then ${\vec L}.{\vec \theta}=L_z \theta$ and
${\vec p}.{\vec \theta}=p_z \theta$. In this situation equation (31)
can be written as
\begin{equation}
r^3 - 2M r^2 - \frac{M}{2} \left[{ L_z \theta} - \frac{1}{8 } \left(
p^2 - p_z^2 \right) \theta^2 \right]+ {\cal O}(\theta^3)+...=0 .
\end{equation}
Since $p^2=p_x^2+p_y^2+p_z^2$, one can write $(p^2 - p_z^2) \theta^2
=(p_x^2+p_y^2) \theta^2$ and therefore (32) can be written as
follows
\begin{equation}
r^3 -2Mr^2 -\frac{M L_z \theta}{2}+ \frac{M}{16} \left( p_x^2 +
p_y^2 \right) \theta^2+ {\cal O}(\theta^3)+...=0 .
\end{equation}
Since Schwarzschild black hole is non-rotating, we set ${\vec L}=0 $
and therefore $L_z=0$( this means that space noncommutativity has no
effect on the Schwarzschild geometry up to first order of space
noncommutativity parameter). So we find
\begin{equation}
r^3 -2Mr^2 + \frac{M}{16} \left( p_x^2 + p_y^2 \right) \theta^2+
{\cal O}(\theta^3)+...=0 .
\end{equation}
With the following definitions
\begin{equation}
a \equiv-2M=-r_s,\;\;\;\eta \equiv \frac{M}{16} \left( p_x^2 + p_y^2
\right) \theta^2 ,
\end{equation}
and considering only terms up to second order of $\theta$, the
radius of event horizon for noncommutative Schwarzschild black
hole becomes
\begin{equation}
{\hat r}_s \equiv \frac{-a}{3}
+ \left(\frac{-2a^3 - 27\eta + \sqrt{108a^3\eta+729\eta^2}}{54}\right)^{1/3}\nonumber\\
+ \frac{a^2}{9}\left(\frac{-2a^3 - 27\eta +
\sqrt{108a^3\eta+729\eta^2}}{54}\right)^{-1/3} .
\end{equation}
Two other roots of (34) are not real. In the case of commutative
space,  $\eta=0$, and therefore we recover usual Schwarzschild
radius, $r_s=2M$. Since $a\gg \eta$, we can expand equation (36) to
find the following relation for Schwarzschild radius in
noncommutative space
\begin{equation}
{\hat r}_s =-a -\frac{\eta}{a^2} -\frac{27}{2}\frac{\eta^2}{a^{5}}.
\end{equation}
Since $a =-r_s$ we have considered only the real parts of our
equations. One can write $\eta=M \alpha$, where
\begin{equation}
\alpha = \frac{1}{16} \left( p_x^2 + p_y^2 \right) \theta^2 ,
\end{equation}
and therefor $\eta=\frac{r_s}{2}\, \alpha$. In this manner, we can
write equation (37) as follows
\begin{equation}
{\hat r}_s =r_s -\frac{\alpha}{2 r_s}
+\frac{27}{8}\frac{\alpha^2}{r_s^{3}}.
\end{equation}
After calculation of Schwarzschild radius of black hole in
noncommutative space, we have all prerequisites to calculate
thermodynamics of black hole in noncommutative spacetime.\\
First we consider black hole temperature. The Hawking temperature of
Schwarzschild black hole in noncommutative space can be given by the
following relation
\begin{equation}
\hat T_H = \frac{M}{2\pi {\hat r}_s {\hat r}_s},
\end{equation}
where substitution of $\hat{r_s}$ leads to the following generalized
statement
\begin{equation}
\hat T_H = \frac{M}{2\pi} \Big(r_s -\frac{\alpha}{2 r_s}
+\frac{27}{8}\frac{\alpha^2}{r_s^{3}}\Big)^{-2}
\end{equation}
which leads to the following relation
\begin{equation}
\hat T_{H}\approx \frac{1}{8\pi M}\Big[1+\frac{\alpha}{4
M^2}-\frac{3 \alpha^2}{8  M^4}\Big].
\end{equation}
Figure 1 shows the plot of black hole temperature versus its horizon
radius in three candidate models. As this figure shows, within GUP
and Noncommutative geometry, black hole before its terminal stage of
evaporation reaches to a maximum temperature and then cools down to
a zero temperature remnant.\\
Now we calculate entropy of black hole in a noncommutative
spacetime. In the standard Bekenstein argument, the relation between
energy and position uncertainty of a given particle is given by(see
[10] and references therein)
\begin{equation}
E \geq \frac{1}{\delta x}  .
\end{equation}
Within a noncommutative framework, we suppose  $\delta x=\hat
r_{s}$. Therefore, we find the following generalization
\begin{equation}
E \geq \frac{1}{\hat r_s},
\end{equation}
which substitution of ${\hat r}_s$ from (39) leads to
\begin {equation}
E \geq \frac{1}{\Bigg(r_s -\frac{\alpha}{2 r_s}
+\frac{27}{8}\frac{\alpha^2}{r_s^{3}}\Bigg)}.
\end{equation}
Since $r_s=2M$, this relation implicitly shows the modification of
standard dispersion relations which has strong support on loop
quantum gravity[9]. In this manner, the increase of event horizon
area is given by
\begin{equation}
\Delta \hat A \geq 4 (\ln 2)\frac{1}{\Bigg(1 -\frac{\alpha}{2
r_s^{2}} +\frac{27}{8}\frac{\alpha^2}{r_s^{4}} \Bigg)}.
\end{equation}
which leads to the following relation
\begin{equation}
 \frac{dS}{d \hat A} \approx \frac {\Delta S_(min)}{\Delta \hat A_(min)}\simeq\frac
{\ln2}{4 (\ln2)\frac{1}{\Bigg(1 -\frac{\alpha}{2 r_s^{2}}
+\frac{27}{8}\frac{\alpha^2}{r_s^{4}}\Bigg)}}.
\end{equation}
Therefore we can write
\begin{equation}
\frac {dS}{d \hat A}\simeq\frac{1}{4}\Bigg[1 -\frac{\alpha}{2
r_s^{2}} +\frac{27}{8}\frac{\alpha^2}{r_s^{4}}\Bigg].
\end{equation}
Now we should calculate $d \hat A$. Since
\begin{equation}
\hat A =4\pi {\hat r}_s {\hat r}_s,
\end{equation}
we find
\begin{equation}
d\hat A =\Bigg[1+\gamma_{1}\Big(\frac{4\pi
\alpha}{A}\Big)^{2}+\gamma_{2}\Big(\frac{4\pi
\alpha}{A}\Big)^{3}+\gamma_{3}\Big(\frac{4\pi
\alpha}{A}\Big)^{4}\Bigg]d A,
\end{equation}
where $\gamma_{i}$'s are some
constant$\Big(\gamma_{1}=-7,\;\;\;\gamma_{2}=\frac{27}{4},\;\;\;\gamma_{3}=-3(\frac{3}{2})^{6}\Big)$
and $A=4\pi r_{s}^{2}$. We can integrate (48) to find
$$S\simeq \frac{A}{4}-\frac{\pi\alpha}{2} \ln{\frac{
A}{4}}+\kappa_{1}(\frac{\pi\alpha}{2})^{2}\Big(\frac{4}{
A}\Big)+\kappa_{2}(\frac{\pi\alpha}{2})^{3}\Big(\frac{4}{
A}\Big)^2+$$
\begin{equation}
\kappa_{3}(\frac{\pi\alpha}{2})^{4}\Big(\frac{4}{
A}\Big)^3+\kappa_{4}(\frac{\pi\alpha}{2})^{5}\Big(\frac{4}{
A}\Big)^4+\kappa_{5}(\frac{\pi\alpha}{2})^{6}\Big(\frac{4}{
A}\Big)^5,
\end{equation}
where $\kappa_{i}$'s are some
constant$\Big(\kappa_{1}=\frac{29}{2},\;\;\;\kappa_{2}=-41,\;\;\;\kappa_{3}=
\frac{1305}{4},\;\;\;\kappa_{4}=-63\times
\Big(\frac{3}{2}\Big)^{4},\;\;\;\kappa_{5}=\frac{3^{10}}{40}\Big)$.
Generally, this relation can be written as
\begin{equation}
S\simeq \frac{A}{4}-\frac{\pi\alpha}{2} \ln{\frac{
A}{4}}+\sum_{n=1}^{\infty}c_{n}\Big(\frac{4}{A}\Big)^{n}+\cal{C}
\end{equation}
Where
\begin{equation}
{\cal{C}}\simeq -\frac{A_p}{4}+\frac{\pi\alpha}{2} \ln{\frac{
A_p}{4}}-\sum_{n=1}^\infty c_{n}\Big(\frac{4}{A_p}\Big)^{n}.
\end{equation}
This is an interesting result which shows the modified entropy of
black hole within noncommutative geometry. In the case of
commutative spaces $\alpha=0$ and this equation yields the
standard Bekenstein entropy,
\begin{equation}
S\simeq \frac{A}{4}.
\end{equation}
Equation (52) is very similar to (20). As a result we see that GUP
and Noncommutative geometry give the same area dependence to the
modified entropy of black hole. This feature may inherently reflect
the fact that GUP and spacetime noncommutativity are not different
in essence. Figure 2 shows the entropy-area relation for an
evaporating black hole in bekenstein-Hawking and the noncommutative
geometry view points. Within noncommutative geometry approach black
hole in its final stage of evaporation reaches to a zero entropy
remnant.
\section{Conclusion}
There are several approaches to incorporate quantum gravitational
effects in thermodynamics of black holes. Here we have developed two
of these approaches with details. In another work[10], we have
calculated quantum correction of black hole thermodynamics using
modified dispersion relations. Our calculations show that overall
behavior (functional form) of entropy-area or temperature-mass
relations are independent of different approaches. For example, we
have shown here that with GUP and Noncommutative geometry NCG one
finds the following relations
$$T_{H}\approx \frac{1}{8\pi M}\Big[1+\frac{\beta}{16
M^2}+\frac{\beta^2}{128  M^4}\Big]\quad\quad in\quad GUP$$
$$\hat T_{H}\approx \frac{1}{8\pi M}\Big[1+\frac{\alpha}{4
M^2}-\frac{3 \alpha^2}{8  M^4}\Big]\quad\quad\quad\quad\quad in\quad
NCG$$ for temperature up to second order of expansion parameter.
These two statements are not different in mass dependence. Similarly
for entropy-area relation we have found
$$S\simeq\frac{A}{4}-\frac{\pi\beta}{4}\ln {\frac
{A}{4}}+\sum_{n=1}^\infty
c_{n}\Big(\frac{4}{A}\Big)^{n}+{\cal{C}}\quad\quad in\quad\quad
GUP$$
$$S\simeq \frac{A}{4}-\frac{\pi\alpha}{2} \ln{\frac{
A}{4}}+\sum_{n=1}^{\infty}c_{n}\Big(\frac{4}{A}\Big)^{n}+{\cal{C}}\quad\quad
in\quad NCG.$$ These similarity may reflect the fact that GUP and
space noncommutativity essentially are not different concepts. In
this way one can obtain easily the relation between GUP and space
noncommutativity parameters by comparison of the corresponding
relations for entropy or temperature. On the other hand, from a
string theory point of view, one can show the following relations
for temperature and entropy of a black hole
$$T=\frac{1}{8 \pi M} \Bigg( 1 - \rho \frac {1}{4M^2}
+ \frac {1}{4M^4} (\rho^2 + \frac{\lambda}{4})\Bigg),$$ and
$$S=\frac{A}{4}+\rho\ln \frac {A}{4} + \lambda(\frac
{4}{A}).$$ These results are more reliable since they are based on
direct analysis of quantum behavior of black hole. Comparison with
previous results shows that GUP and space noncommutativity are
two similar concept of string theory.\\
In addition, our analysis shows that space noncommutativity has no
effect on the structure of a Schwarzschild black hole in the first
order approximation in noncommutativity parameter. The final stage
of black hole evaporation is a remnant with absolute zero
temperature and also zero entropy. Before black hole reach its final
state as a remnant, it reaches to a maximum temperature and then
cools down to zero temperature. A direct calculation of curvature
shows that this remnant is not singular[11]. Note that our works
differs from previous findings in two main respects: previous works
based on GUP have considered final stage of black hole as a remnant
with non-zero temperature(see for example[5,6,7]). Here we have
shown that actually this remnants should have zero temperature.
Secondly, our approach based on space noncommutativity differs from
existing works such as [11,12] since we have not used the smeared
picture of objects with Gaussian profile as a result of space
noncommutativity, instead we have considered direct generalization
of Schwarzschild radius in a perturbational framework. In other
words, we have considered the effect of noncommutativity on
geometric part of Einstein's equations whereas Nicolini {\it et al}
have considered the effects of space noncommutativity on matter part
of Einstein's equations. Although these approaches seems to be
different in their view point on space noncommutativity, but their
results are similar in many respects. one can easily see the close
coincidence of these two approaches by comparing for example the
temperature of black hole calculated in these two view points; our
figure $1$  is in complete agreement with figure $4$ of Nicolini
{\it et al} first paper in reference[11].

\begin{figure}[ht]
\begin{center}
\includegraphics{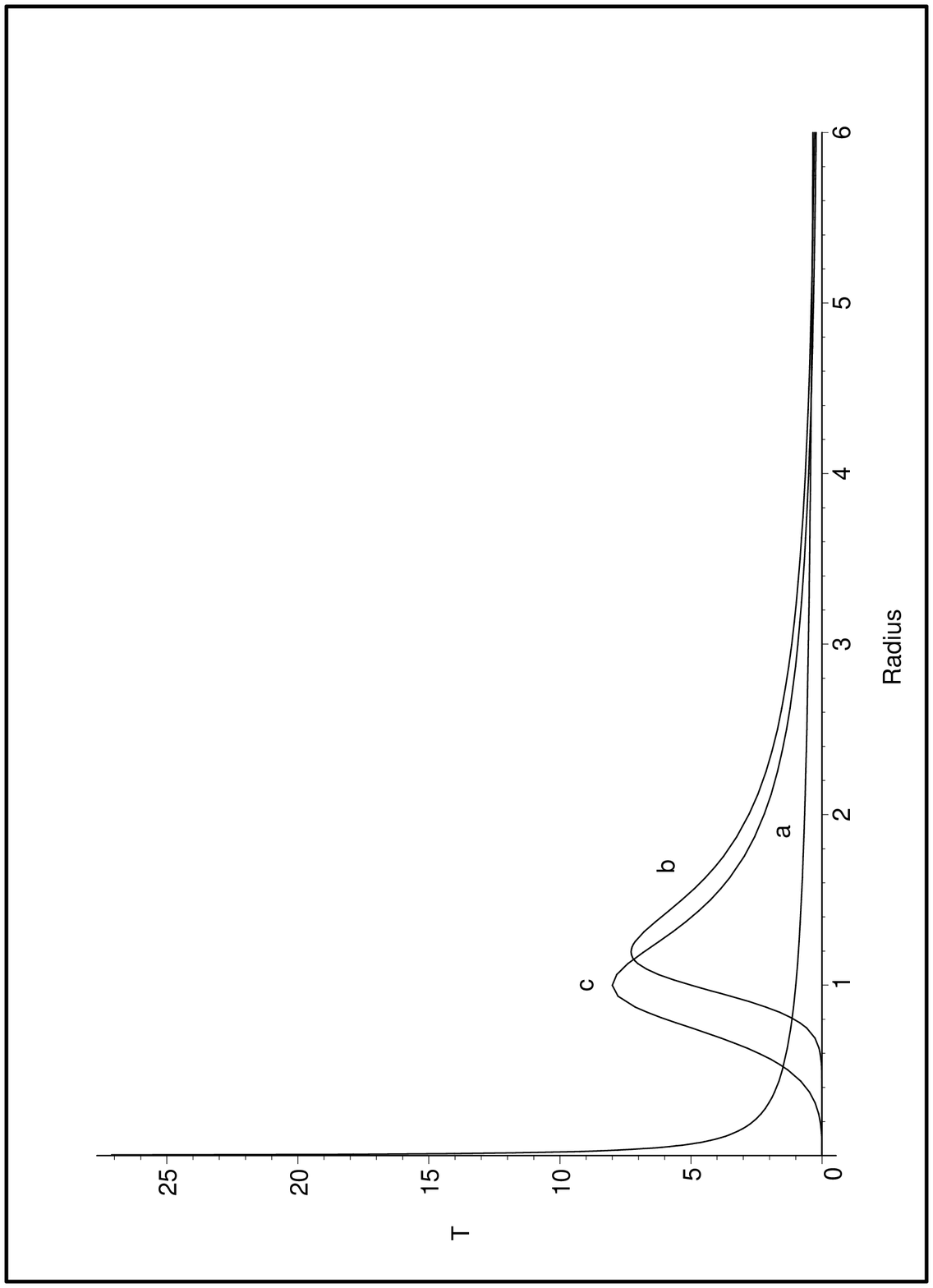}
\end{center}
\vspace{16 cm}
 \caption{\small {Black hole temperature versus its radius of event horizon in
 three candidate models: a) Bekenstein-Hawking Model, b) NCG  and c) GUP . }}
 \label{fig:1}
\end{figure}

\begin{figure}[ht]
\begin{center}
\includegraphics{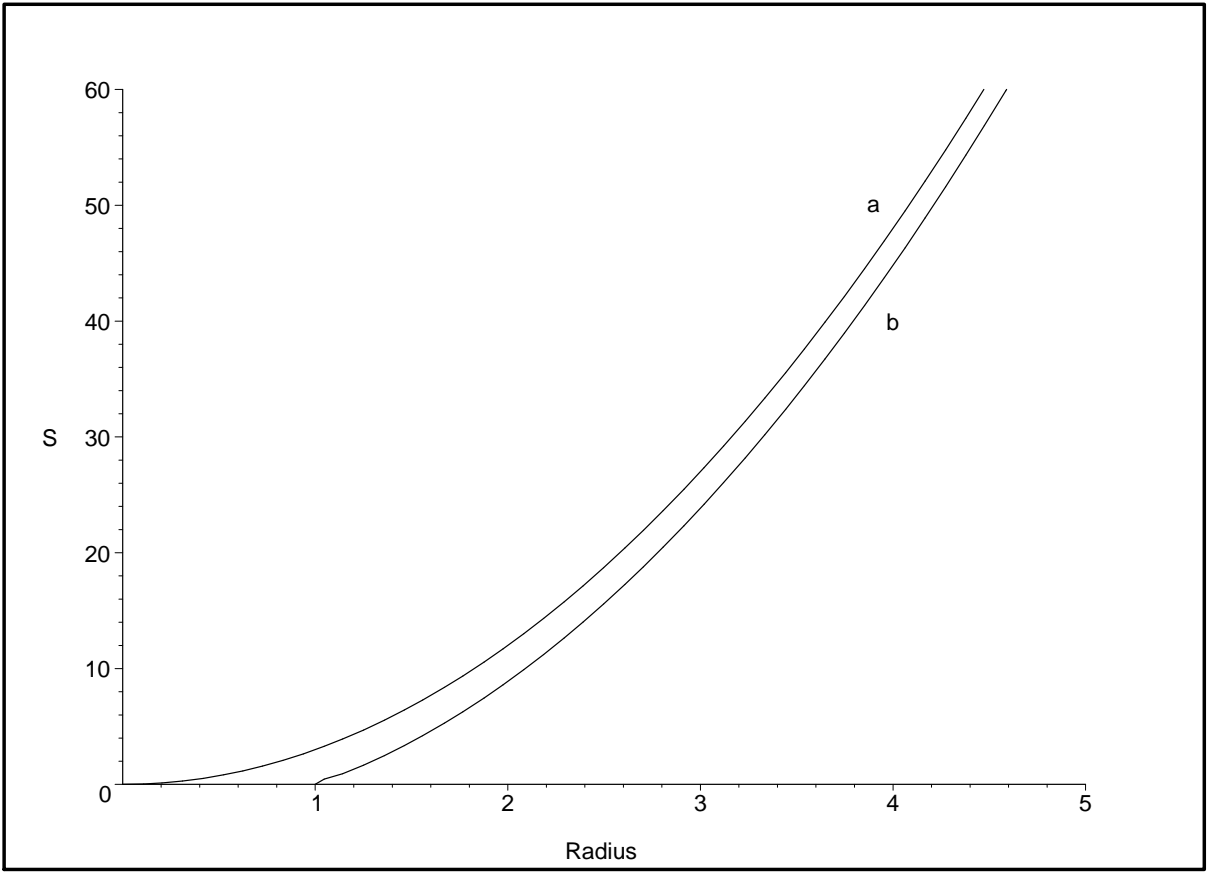}
\end{center}
\vspace{16 cm}
 \caption{\small {Black hole entropy versus its radius of event horizon in
two candidate models: a) Bekenstein-Hawking Model, b) NCG . GUP
result is similar to curve b.}}
 \label{fig:1}
\end{figure}

\end{document}